\documentclass{ws-procs9x6}

\def\pn{\par\noindent} 
\def\gsimeq{\mathrel{\hbox{\rlap{\hbox{\lower4pt\hbox{$\sim$}}}\hbox{$>$}}}}
\def\lsimeq{\mathrel{\hbox{\rlap{\hbox{\lower4pt\hbox{$\sim$}}}\hbox{$<$}}}}
\def\cgs{${\rm erg ~cm^{-2} ~s^{-1}}$}
\def\xmm{XMM--{\it Newton}}

\newcommand{\aj}[2]{\mbox{ {\em A.J.\ }{\bf #1}, {#2}}}
\newcommand{\apjs}[2]{\mbox{ {\em Ap.J.S.S.\ }{\bf #1}, {#2}}}
\newcommand{\aap}[2]{\mbox{ {\em A.\&A.\ }{\bf #1}, {#2}}}
\newcommand{\mnras}[2]{\mbox{ {\em M.N.R.A.S.\ }{\bf #1}, {#2}}}

\begin{document}

\title{Are QSO2 Hiding among EROs?}

\author{M. BRUSA\footnote{\uppercase{T}he results presented at this
		conference have been obtained in collaboration with
		\uppercase{A}ndrea \uppercase{C}omastri,
		\uppercase{E}manuele \uppercase{D}addi,
		\uppercase{L}ucia \uppercase{P}ozzetti,
		\uppercase{G}ianni \uppercase{Z}amorani,
		\uppercase{A}ndrea \uppercase{C}imatti,
		\uppercase{C}ristian \uppercase{V}ignali,
		\uppercase{F}abrizio \uppercase{F}iore, and
		\uppercase{M}arco \uppercase{M}ignoli. }} 
 
\address{Dipartimento di Astronomia Universit\`a di Bologna \& \\
INAF--Osservatorio Astronomico di Bologna, \\
via Ranzani, 1 \\
I--40127 Bologna \\
E-mail: brusa@bo.astro.it}


\maketitle

\abstracts{
We present the results of a deep (80 ks) XMM-{\it Newton} 
survey of the largest sample of near-infrared selected Extremely Red
Objects (R-K $>$ 5) available to date ($\sim 300$ objects\cite{d00}).
The fraction of individually detected, X--ray emitting EROs
is of the order of $\sim 3.5$\%, down to F$_x\gsimeq 4\times 10^{-15}$
cgs and K$s<19.2$.
In order to derive the X--ray intrinsic properties of AGN EROs 
and to place our findings in a broader context, we have also considered all
the X-ray detected EROs 
available in the  literature.
The X-ray, optical, and near-infrared properties of those X-ray
selected EROs with a spectroscopic or photometric redshift 
nicely match those expected for quasars 2, the
high-luminosity, high-redshift obscured AGNs predicted in XRB
synthesis models.}

\section{Introduction}
The hard X--ray selection turned out to be very efficient 
in revealing an AGN population with optical to near--infrared colours 
redder than those of optically selected QSOs.
In this respect, the discovery that a sizable fraction of hard X--ray sources
also associated to extremely red objects (EROs) with optical to
near--infrared R-K $>$ 5 colour is even more 
intriguing\cite{lehmann,mainieri,willott}.  
Given the key role played by EROs in the cosmological scenario, 
hard X--ray observations can help to constrain the fraction of AGN
among the ERO population and, at the same time, to provide an exciting 
opportunity  to investigate the link between nuclear activity and galaxy
formation\cite{granato}. 

\subsection{Fraction of AGN EROS}
In this framework, we have started an extensive program of 
multiwavelength observations of the 
largest sample of EROs available to date\cite{d00}, 
selected in a contiguous area ($\sim 700$ arcmin$^2$)
down to a magnitude limit of K$s=$19.2.
We have obtained a total of $\sim$80 ks observation with \xmm; 
the high--energy throughput of \xmm, coupled with the large field 
of view, are well--suited to assess the fraction of AGN among 
a statistically significant sample of EROs 
at relatively bright X--ray fluxes\cite{brusa04}. \\
Among the 257 EROs which fall within the \xmm\ area ($\sim 380$
arcmin$^2$) analysed in 
Brusa et al. (2004), nine are individually detected in the X--rays.
The fraction of X--ray detected (i.e. AGN--powered) EROs at 
K$s$=19.2 and F$_{0.5-10 keV}\gsimeq 2\times 10^{-15}$ \cgs is therefore 
$\sim$ 3.5\%. Conversely, the fraction of EROs among hard X--ray sources
is much higher ($\sim$15\%).
%
\subsection{X--ray to optical properties}
In order to investigate the nature of hard X--ray selected EROs
and the link between faint hard X--ray sources and the 
ERO population, we have collected from the literature a sample of 
118 X--ray detected EROs (including 9 in our sample);
for 52/118 photometric or spectroscopic redshifts 
are available (data 
from Lockman Hole\cite{mainieri}, 
CDFN\cite{barger}, CDFS\cite{szokoly}, literature\cite{willott,crawford,brusa}).
This sample is by no means homogeneous (e.g. the selection criteria
for EROs are slightly different, R-K$>5$ or I-K$>4$ depending on the
authors; or the K--coverage is not complete), but 
could be considered representative of EROs individually
detected in the X--rays. \\
   \begin{figure*}[!t]
\vspace{-1.1cm}
{\par
  {\hbox to \hsize{%
	\psfig{file=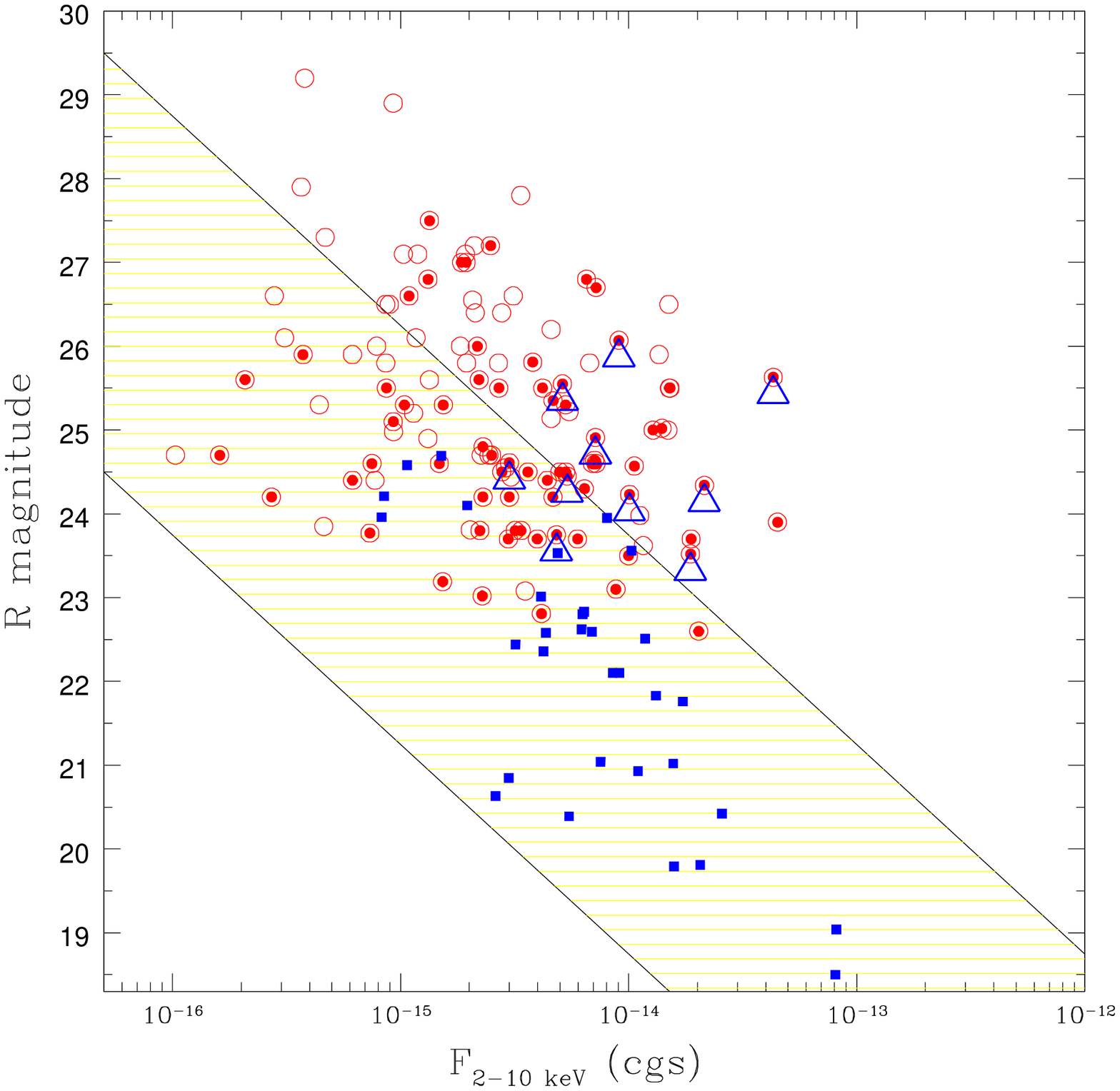,width=2.25in}
	\psfig{file=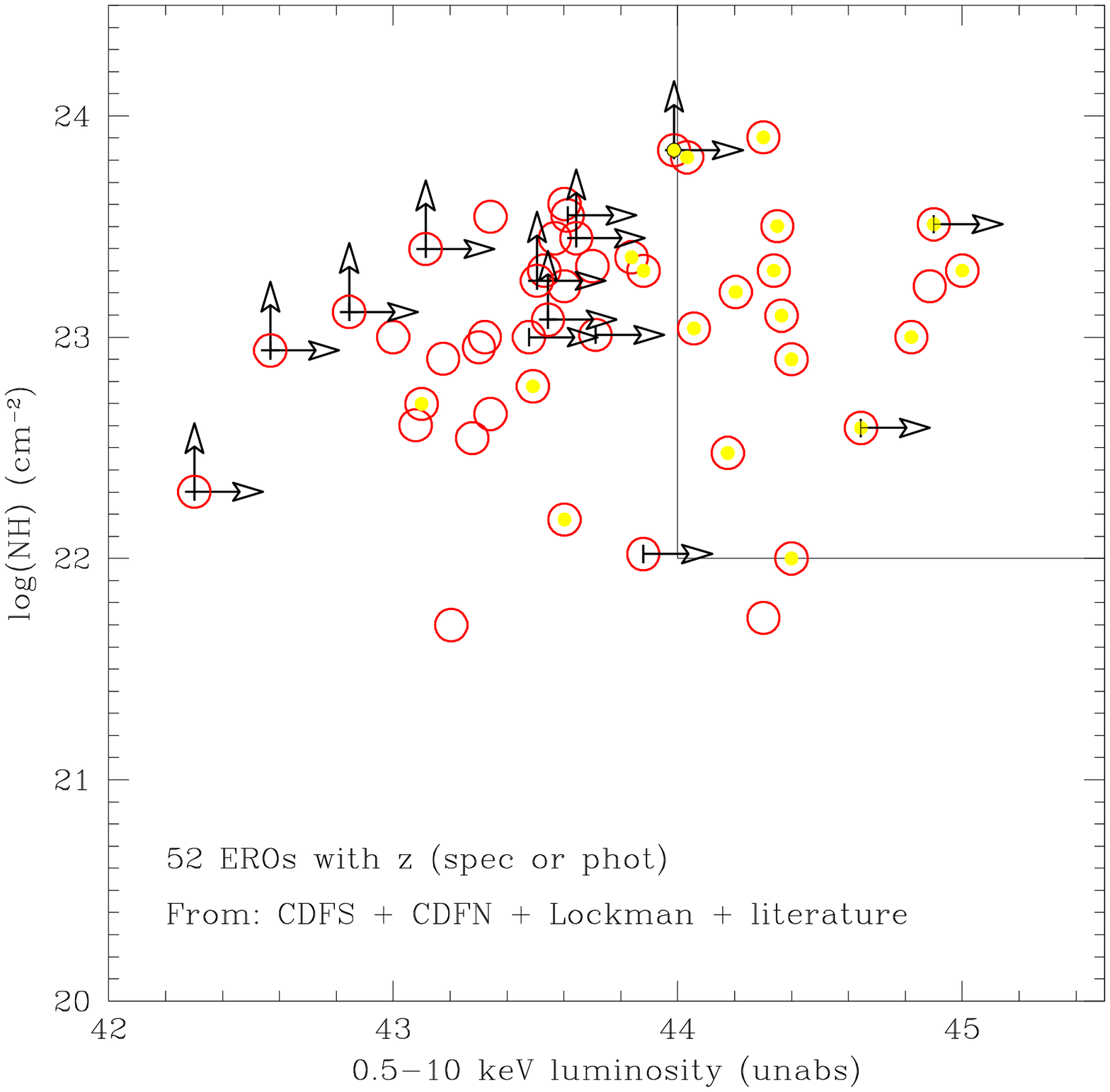,width=2.25in}}}}
   \caption{({\it left panel})
 R--band  magnitude vs. hard X--ray flux
for EROs, serendipitously detected in hard X--ray surveys. (Triangles = EROs in the ``Daddi Field''; 
Open circles = EROs in the reference sample;
filled circles = EROs with redshifts -- see text for details).
Broad Line AGN detected in the CDFS and CDFN surveys 
are also reported (small squares).
The shaded area represents the region occupied by known AGN
(e.g. quasars, Seyferts, emission line galaxies) along the correlation
log$(X/O)=0\pm 1$.
({\it right panel}) Logarithm of the unabsorbed, full band X--ray
  luminosity  versus 
the logarithm of the absorbing column density (N$_H$) for all the
X--ray detected EROs with spectroscopic or photometric redshifts from
the comparison sample (open circles). Filled symbols are EROs with X/O $>10$
(see text). The boxy region indicates the locus of QSO2.}
    \label{hard}
    \end{figure*}
The R--band magnitudes plotted versus the hard X--ray fluxes are
reported in Fig.~\ref{hard} (left panel): 
about half of the sources show an X--ray--to--optical flux
ratio (X/O) larger than 10, shifted up by one order of magnitude from
that of BL AGN, confirming independent results from near 
infrared observations of X--ray sources selected on the basis of their 
high X/O\cite{mignoli}.

\section{X--ray Properties of AGN EROs}
First results suggested that the AGN population
among EROs, although not dominant, 
shares the same X--ray properties of high luminosity, highly
obscured AGN, the so--called quasar 2 (QSO2)\cite{mainieri,alexander,severgnini}.\\
In order to check whether X-ray absorption is common among these 
objects, we have quantitatively estimated the intrinsic X--ray
column densities for the 52 EROs with a reliable spectroscopic or
photometric identification.\\
Column densities for the sources detected in the CDFN and CDFS have
been obtained by fitting the observed counts with a single power law
model plus absorption in the source rest--frame.  For the sources in
the Lockman Hole and in the ``Literature'' sample, the best--fit values
quoted by the authors have been adopted. In all
the cases, X--ray luminosities were estimated from the observed X--ray
fluxes and corrected for absorption.
The results are reported in the right panel of Fig.~\ref{hard}.  \\ 
Almost all of the individually detected EROs have
intrinsic N$_H>10^{22}$ cm$^{-2}$, and they
actually {\it are} heavily obscured AGN. 
This study 
confirms previous evidences mainly based on a
Hardness Ratio analysis\cite{alexander} and on few isolated 
examples\cite{crawford,willott,stevens}, and unambiguously indicates that large columns of cold
gas (even $> 10^{23}$ cm$^{-2}$) are the rule rather than the
exception among X--ray bright EROs.
%
\section{EROs and QSO2: a selection criterion}
\pn Given the high--redshift of these objects (z$\gsimeq 1$) and the
average X--ray flux of the comparison sample ($\sim 4\times 10^{-15}$
\cgs\ ), it is not surprising that the majority of X--ray detected
EROs have high X--ray luminosities (L$_X>10^{43}$ erg s$^{-1}$). 
Moreover, according to our analysis, a significant fraction
of them have X--ray luminosities exceeding 
$10^{44}$ erg s$^{-1}$, and therefore lie within the quasar
regime.  
The large intrinsic column densities further imply that AGN
EROs, selected at the brightest X--ray fluxes, have properties similar
to those of QSO2, the high--luminosity, high--redshift Type 2
AGNs predicted by X--Ray Background synthesis models\cite{comastri,gilli}.
Among the X--ray detected EROs, the higher is the luminosity, the
higher is the X--ray to optical flux ratio (filled symbols 
in right panel of Fig.~\ref{hard}). This confirms that a
selection based on X/O $>10$ is a powerful tool to detect
high--luminosity, highly obscured sources, 
and it is even stronger when coupled with a previous selection on the
extremely red colors. \\
Given that the search for QSO2
on the basis of detection of narrow optical emission lines is very
difficult and is already challenging the capabilities of the largest
optical telescopes, the proposed ``alternative'' method which combines
near--infrared and X--ray observations, could provide a
powerful tool to uncover luminous, obscured
quasars. 

\section*{Acknowledgments}
I kindly acknowledge support by INAOE, Mexico, during the
2003 Guillermo-Haro Workshop where part of this work was performed.

\end{document}